# Deep learning mediated single time-point image-based prediction of embryo developmental outcome at the cleavage stage

Manoj Kumar Kanakasabapathy[1]†, Prudhvi Thirumalaraju[1]†, Charles L Bormann[2,3]†, Raghav Gupta[1], Rohan Pooniwala[1], Hemanth Kandula[1], Irene Souter[2,3], Irene Dimitriadis[2,3], Hadi Shafiee[1,3]*

[1] Division of Engineering in Medicine, Department of Medicine, Brigham and Women's Hospital, Harvard Medical School, Boston, Massachusetts, United States of America

[2] Division of Reproductive Endocrinology and Infertility, Department of Obstetrics & Gynecology, Massachusetts General Hospital, Harvard Medical School, Boston, Massachusetts, United States of America

[3] Department of Medicine, Harvard Medical School, Boston, Massachusetts, United States of America

*Corresponding author: Hadi Shafiee

E-mail: hshafiee@bwh.harvard.edu

† These authors contributed equally to this work

## Abstract

In conventional clinical in-vitro fertilization practices embryos are transferred either at the cleavage or blastocyst stages of development. Cleavage stage transfers, particularly, are beneficial for patients with relatively poor prognosis and at fertility centers in resource-limited settings where there is a higher chance of developmental failure in embryos in-vitro. However, one of the major limitations of embryo selections at the cleavage stage is the availability of very low number of manually discernable features to predict developmental outcomes. Although, time-lapse imaging systems have been proposed as possible solutions, they are cost-prohibitive and require bulky and expensive hardware, and labor-intensive. Advances in convolutional neural networks (CNNs) have been utilized to provide accurate classifications across many medical and non-medical object categories. Here, we report an automated system for classification and selection of human embryos at the cleavage stage using a trained CNN combined with a genetic algorithm. The system selected the cleavage stage embryo at 70 hours post insemination (hpi) that ultimately developed into top-quality blastocyst at 70 hpi with 64% accuracy, outperforming the abilities of embryologists in identifying embryos with the highest developmental potential. Such systems can have a significant impact on IVF procedures by empowering embryologists for accurate and consistent embryo assessment in both resource-poor and resource-rich settings.

# 1. Introduction

In most fertility clinics, embryo transfers are performed at cleavage or blastocyst stage of development. Embryos are at the cleavage stage 2-3 days after fertilization and reach the blastocyst stage 5-7 days after fertilization. Cleavage stage embryos are generally selected for transfer based on cell number, degree of cellular fragmentation, and the overall symmetry of the blastomeres (Prados et al., 2012). Traditional methods of embryo selection rely on visual embryo morphological assessment and are highly practice-dependent and subjective. Advances in time-lapse imaging (TLI) techniques have enabled regular and automated data acquisition of embryo development under controlled environments, along with identifying objective morphokinetic parameters (Hlinka et al., 2012; Cruz et al., 2012; Lechniak et al., 2008; Lemmen et al., 2008; Azzarello et al., 2012). However, the developed time-lapse embryo selection algorithms (ESAs) have shown promising results only in identifying embryos with low developmental potential when used by trained embryologists and the addition of TLI systems to conventional manual embryo testing did not improve clinical outcomes but increased the embryo morphology assessment times (Conaghan et al., 2013; Kirkegaard et al., 2015; Kaser D.J, 2016; M. Chen et al., 2017). Other studies have also shown that the addition of time-lapse imaging systems to conventional manual embryo testing not only did not improve clinical outcomes but also increased the time of the embryo assessment (Kaser D.J, 2016; M. Chen et al., 2017). Potentially owing to these reasons and their high principal costs, most fertility centers do not own or operate such systems and are limited to manual evaluations of embryos (Dolinko et al., 2017).

While the most common embryo selection methods rely on embryo morphological appearance, alternative methods involve genotypic and phenotypic analyses (L. Richter, 2008). The methods involving genetic analysis have shown promise in identifying embryos capable of achieving successful pregnancies, however, they are invasive and have raised concerns about the potential damage caused to embryos due to the biopsy procedure (Dahdouh et al., 2015; Brezina et al., 2016; Minghao Chen et al., 2015; Gleicher and Orvieto, 2017). Similarly, non-invasive methods such as the analysis of metabolites and proteins in culture medium have also been explored for embryo selection. However, these methods are still limited by the need for expensive and sensitive instruments since they are dependent on samples with limited analytes. Overall, alternative methods of embryo analysis are either not cost-effective or invasive and require further validation of their effectiveness (L. Richter, 2008; Uyar and Seli, 2014; Gleicher and Orvieto, 2017). Therefore, static traditional morphological assessments are widely used owing to their simplicity and proven cost-effectiveness relative to the alternative methods (L. Richter, 2008). Here, we employ a deep-convolutional neural network (CNN) to predict embryonic development and combined with a genetic algorithm to select the top-quality embryo (Fig 1).

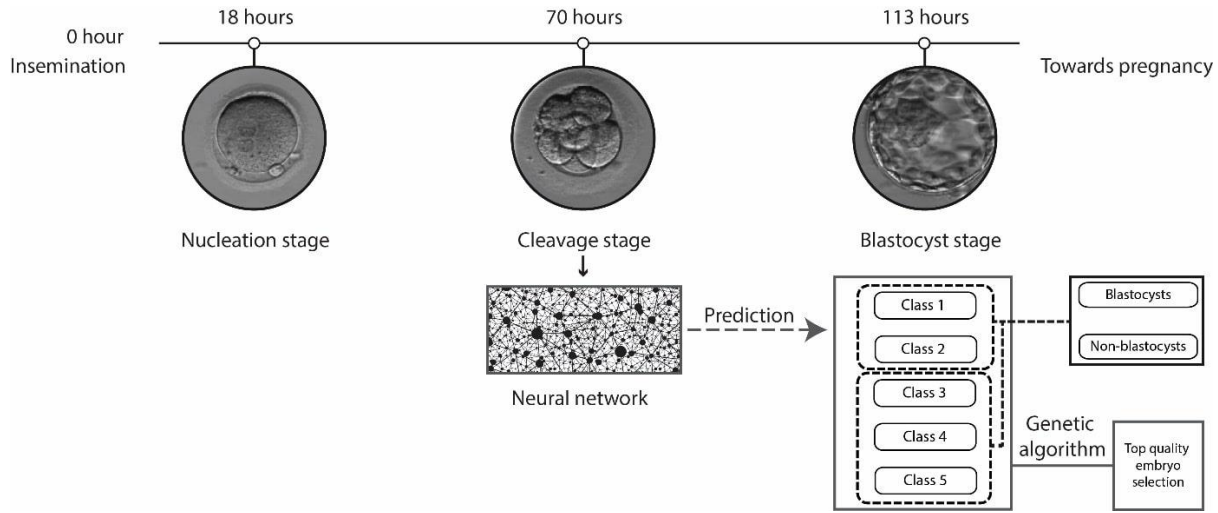

**Fig 1. The artificial intelligence (AI)-based embryo evaluation system.**

Images of patient embryos were extracted to obtain the embryo morphologies at time points 70, and 113 hours post-insemination (hpi). The extracted data is passed through the pre-trained neural network scheme based on the Xception architecture. The network assesses and classifies the embryos based on its morphological features and when combined with the developed genetic algorithm the system selects up to two embryos of the highest quality.

Deep convolutional neural networks, used in this work, does not use hand-crafted features and processes non-segmented raw images as input. The networks select the best set of features through iterative learning cycles at the pixel-level directly from images unlike prior approaches involving the use of machine learning and pattern recognition techniques that utilized human-designed features (Manna et al., 2013; Patrizi et al., 2004; Morales et al., 2008). The trained neural network evaluated embryo images at pre-blastocyst stage and predicted their eventual development into blastocysts by identifying features of pre-blastocyst embryo morphology that are associated with eventual blastocyst formation.

Embryos were analyzed at 70 hours post insemination (hpi) and were matched to their outcomes at 113 hpi. These embryos were categorized into 5 classes based on their eventual quality of development at 113 hpi (Fig 2). The two major categories of non-blastocysts and blastocysts included the training classes 1, 2, and 3, 4, 5, respectively.

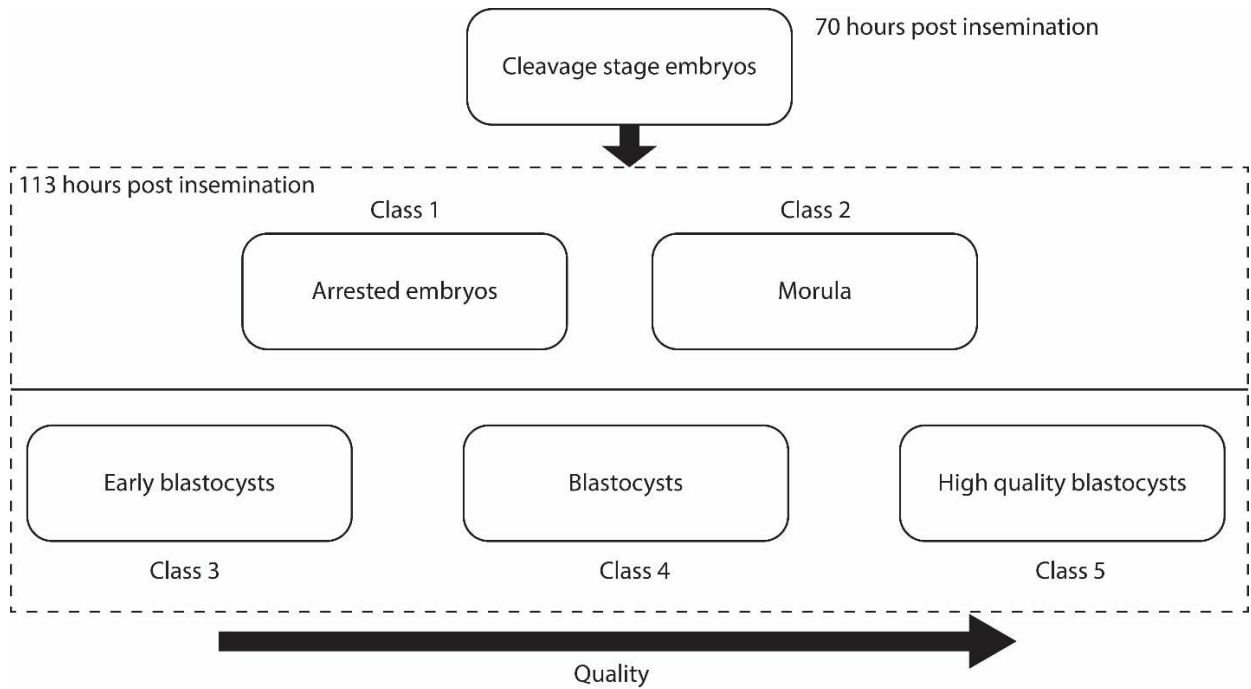

**Fig 2. Embryo hierarchy used by the neural network.**

Following insemination, pronuclear stage embryos are categorized into two classes and based on their 113-hours morphologies are sorted into 2 major classes (blastocysts and non-blastocysts) subdivided into 5 classes. Embryos with abnormal fertilization (non-2PN) were not tracked further and thus were not considered in 70 hpi and 113 hpi assessments. Classes 1 and 2 were composed of non-blastocysts and classes 3, 4, and 5 were composed of blastocysts. Class 5 composed of blastocysts that met the clinical criteria for cryopreservation.

Using a retrospective dataset comprising of 2449 embryos, the deep CNN model was trained and tested to primarily predict the developmental outcome between two classes (non-blastocysts and blastocysts defined at 113 hpi) using images of cleavage stage embryos captured at 70 hpi. Furthermore, we have also tested the network in combination with a genetic algorithm in selecting top quality embryos (embryos that developed into high quality blastocysts) from a cohort. Through testing a set of embryos collected from 100 patients, the CNN performed with 70.05% accuracy in assessing embryos at 70 hpi. Importantly, such a network when combined with a genetic algorithm can perform embryo selections outperforming highly-trained embryologists at 70 hpi when tested using 97 patient embryo cohorts.

## 2. Materials and Methods

### 2.1 Data collection and preparation

Data was collected at the Massachusetts General Hospital (MGH) fertility center in Boston, Massachusetts. We used 3,469 recorded videos of embryos collected from 543 patients under an institutional review board approval (IRB#2017P001339). Videos were fragmented to extract the frames linked to specific time points (70 hpi) using a custom python script, which made use of the OpenCV and Tesseract libraries. Machine generated timestamps available on each frame of the video was used to identify the images associated to the specific time points. All embryos used in the study were annotated using images from the fixed time-points by senior level embryologists with a minimum of 5 years of human IVF training. Out-of-focus images were included in the datasets and used for both testing and training. Only images of embryos that were completely non-discernable were removed from the study as part of the data cleaning procedure.

## 2.2 Data organization and hierarchical structuring

Only embryos with normal fertilization were used for evaluations at 70 hpi and 113 hpi. The embryo images at 70 hpi and 113 hpi time points were categorized between training classes 1 through 5 (Fig 2). The embryo class categorizations were based on the embryos' eventual developmental state achieved by 113 hpi. Class 1 comprised of degenerated and arrested embryos, which did not begin compaction while class 2 comprised of embryos that were at the morula stage at 113 hpi. Classes 1 and 2 together formed the inference class of 'non-blastocysts'. Class 3 comprised of embryos exhibiting features of an early blastocyst such as the presence of a blastocoel cavity and a thick zona pellucida with lack of overall embryo expansion. Class 4 was made up of embryos, which were blastocysts with blastocoel cavities occupying over half of the embryo volume and possessed either poor inner cell mass (ICM) or poor trophectoderm (TE). These embryos were overall considered to fall below 113 hpi cryopreservation quality criteria based on the MGH fertility center guidelines ($> 3CC$), where 3 represents the degree of expansion (range 1-6) and C represents the quality of ICM and TE (range A-D), respectively. Class 5 on the other hand comprised of all embryos, which met cryopreservation criteria and included full blastocysts to hatched blastocysts. Classes 3, 4, and 5 together formed the inference class of 'blastocysts' that was used in this study.

## 2.3 Neural network development and evaluation

The Xception architecture pre-trained with 1.4 million images of ImageNet was used, which performed with a top-1 accuracy of 79% and top-5 accuracy of 94.5% across 1,000 classes of ImageNet database and fine-tuned the pre-training weights across all layers through transfer learning to fit our dataset and differentiate across the categories of embryos by recognizing relevant features. During the transfer learning process, we discarded the last fully connected layer of the original network and added a new fully connected layer, which classifies the features into the defined five categories. 70 hpi evaluation dataset included images of 2,449 embryos categorized across five classes post-cleaning, based on their clinical annotations made at 113 hpi. The training set comprised of 1,190 images (class1-1.16: class2-1.03: class3-1.19: class4-1: class5-1.5) with a validation dataset of 511 images (1.29: 1.18: 1.41: 1: 1.82). Our independent test set of non-overlapping images was composed of 748 (1.59: 1.20: 1.45: 1: 3.16) cleavage stage embryo images annotated based on their developmental fate at 113 hpi. Whereas the 113 hpi evaluation dataset included images of 2,440 embryos categorized across five classes post-cleaning based on their clinical annotations made at 113 hpi. Our training set for this classification task used 1,188 images (1.14: 1.04: 1.19: 1: 1.53) with a validation dataset of 510 images (1.23: 1.18: 1.31: 1: 1.65) obtained at 113 hpi. The independent non-overlapping test set consisted 742 images (1.69: 1.26: 1.52: 1: 3.26). With the availability of unskewed validation sets prior to augmentation, we used a data generator during training, which performed random rotations and flips across all classes on the fly. For classification at the inference level, the algorithm outputs five confidence values mapping the probabilities of the tested embryo associated with each of the five training classes. The embryo is categorized into the class with the highest confidence.

For comparisons of the neural network's predictive performance with embryologists, retrospectively collected stemming from across 8 embryologists was used. The data was collected during routine clinical evaluations using 44 and 70hpi images, and time-lapse information such as P2 (duration of the 2-cell stage) and P3 (duration of the three-cell stage). A total of 803 normally fertilized embryos whose blastocyst status was known were evaluated (Table S2).

## 2.4 Embryo selection algorithm and testing

To select the top-quality embryo for 70 hpi transfers, we first classified embryos using the standard classification algorithm then scored each embryo by obtaining a weighted sum of the five class confidence values. The weights were optimized through a genetic algorithm trained to select the embryo of the highest quality using the validation data sets used in evaluation models. The accuracy of the selection is established by comparing the class of the highest scored embryo in a patient embryo cohort to the highest class of embryos determined through manual analysis. Selections were performed at two levels: (i) at the inference class level (between blastocysts and non-blastocysts), and (ii) at the training class level (between the five classes). For any given patient set of embryos that are tested with the network, the CNN outputs the probability of each embryo belonging to one of the 5 training classes. However, clinically, a continuous metric is preferred for ranking embryos towards establishing the order of transfer. In order to use the five probability values effectively for calculating the embryo rank, we utilized a genetic algorithm, which is well-suited for optimization problems with multiple existing solutions.

During testing with the dataset of 97 patients, a double-blinded study was performed where the CNN was evaluated against embryologists. From the common test dataset of 100 patients, we removed 3 patients due to their failure to meet the selection criteria. The embryologists selected the best embryos within the patient cohort using a custom designed iOS application specifically developed for obtaining their selections (S1 Fig). The embryologists selected the top quality embryos at 70 hpi using an embryo selection method described previously (Dimitriadis et al., 2017). For accuracy measurements, the selections made by the AI system and embryologists were compared to the historical embryo annotations categorized into the 5 training and 2 inference classes. If the selected top-quality embryo was an embryo of the highest determined embryo class based on historical annotation within a patient embryo cohort, the selection was marked as correct, otherwise marked as incorrect. Both the embryologists and the system were tested in selecting embryos when one (SET) and two (DET) embryo selection/s were made. For DET, the accuracy measurements were calculated with the criteria that at least one selected embryo belongs to the highest class determined through historical annotation within a patient embryo cohort for the performed level of analysis (2 class level or 5 class level).

## 2.5 Genetic algorithm

We trained a genetic algorithm to select the morphologically highest quality embryo from a given cohort. The CNN outputs five probability values in defining an embryo's quality, however a continuous metric is preferred for ranking embryos towards establishing the order of transfer. In order to use the five probability values generated by the CNN for effectively calculating the embryo rank, we utilized a genetic algorithm. This genetic algorithm has four phases starting with initialization, selection, crossover, and mutation.

Embryos from the training and validation data were evaluated with the trained neural network. The list of network classified embryos for each patient was used by the genetic algorithm. The confidence values were combined the weights generated by the genetic algorithm and was the weights optimized such that the best quality embryos were ranked the highest for each patient set within the training data.

An initial population of weights was generated at random. A population size of 100 was initialized where each weight was a 5×1 matrix that represents one of the possible solutions towards the identification of a top-quality embryo using the 5 quality-based training classes. These values were used in the calculation of the fitness, which used the dot product of the weights with the output logits provided by the CNN. The algorithm runs through multiple iterations to select the optimized weights towards the correct identification of the embryo with the highest morphological quality. In each iteration, all the weights of the given population were used to select embryos for each patient (cohort of embryos) within the training set. The top 20 weights with the highest scores were selected in every iteration.

These selected weights (specimens) were then bred with each other with a probability of 20%. It randomly selected 2 specimens from the selected top pool and created a random binary 5×1 matrix, where 1 represents that the given element should be switched in cohort and 0 represents that given element should not be switched within the cohort. The fitness function checks if the selected embryo belongs to the highest class available within the tested cohort. It checks if the selected solution (specimen) picked the embryo belonging to the top class in a given cohort of patient embryos. If the selected embryo belonged to the top class, the score was increased and if it did not, the score was not modified. After iterating for all patients' cohorts, the total scores were used to select the best 20 weights of the given population and were taken for crossover and mutation to repeat the process. The new specimens replaced their parents in the top selected group of embryos. Otherwise the matrix remained the same.

After breeding, each specimen from the top selected group was mutated to give 5 mutations by adding a random float 5×1 matrix with a probability of 20%. These mutations were then added to the new population and the selection step was repeated with the new population of 100. The genetic algorithm ran until the entire population converged to the same score (the maximum score for the given CNN model) after which a random weight was selected from the population as the final weight. Thus, final generated weights were used to further test the embryo cohorts within our test set.

## 2.6 Data visualization techniques

Keras vis environment was used for data visualization. Saliency mapping is a type of image segmentation, that is commonly used to highlight pixels based on their importance in images. Saliency maps were used in our study to visualize the pixels used in the network's decision-making process differentiating them based on their importance. We mapped the activations of the activation layer prior to bottleneck. We used 70 hpi test set images in the generation of the saliency maps. t-distributed stochastic neighbor embedding (t-SNE) was performed to observe the distribution of the test dataset and verify if the CNN was able to isolate embryos into clusters based on their features. t-SNE was used to generate distribution over pairs of embryo images in such a way that similar embryos are grouped closer together while dissimilar embryos are pushed apart. We used the fully connected layer after global average pooling which has 2,048 dimensional vectors in visualizing the similarities between the embryo images, as understood by the trained network, using the respective datasets. Initially, a principal component analysis (PCA) was performed to reduce 2,048 dimensions to 50 and then t-SNE was performed to reduce the 50 dimensions into 2 dimensions for visualization. We have utilized PCA for the initial dimensionality reduction to 50 from 2,048 dimensions, since it helps in suppressing noise while improving computational speed (van der Maaten and Hinton, 2008).

## 2.7 Statistical analysis

Receiver operating characteristics (ROC) analyses were performed using MedCalc 14.8.1. Classification accuracies, sensitivity, and specificity were calculated using both Microsoft Excel and MedCalc. The 95% confidence intervals are Clopper-Pearson binomial confidence intervals. One sample t-tests and "N-1" Chi-squared tests were performed to compare accuracies of embryologists and the network. The significance values for the statistical analyses used, was set at 5%.

# 3. Results

## 3.1 Day 3 Embryo Blastocyst Prediction

Transferring cleavage stage embryos on the second or third day of embryo development is a common procedure in reproductive medicine and is performed by many fertility centers across the world. The CNN was trained and validated for embryo assessment at 70 hpi, based on embryo developmental outcomes at 113 hpi. The algorithm was trained to make developmental predictions at 70 hpi across five classes (Fig 2), which were then consolidated into two major inference categories (blastocysts and non-blastocysts). The embryo data annotation based on morphological blastocyst status was performed at 113 hpi. Annotated embryo images recorded at 70 hpi (n=1,190) and 113 hpi (n=1,190) were used to train the CNN for blastocyst prediction at 70 hpi. Using a test set of 748 embryos, the accuracy of the algorithm in predicting blastocyst development at 70 hpi was 71.87% (CI: 68.41% to 75.15%) (Fig 3). Visualization through t-SNE and saliency revealed that although the system utilizes embryo specific features, the network is experiencing a less than ideal degree of confusion in separation of classes (S2 Fig and Fig 3A). The micro-average and macro-average AUC for the 5 classes based on the ROC analysis were 0.6601 and 0.6096, respectively (S3A Fig). The AUC values for classes 1, 2, 3, 4, and 5 were 0.7677, 0.5429, 0.5101, 0.5802, and 0.6429, respectively (S3A Fig). Confusion matrices mapping the CNN's performance with the test set, in separating between the five training classes showed confusion between classes 2, 3, and 4 (S4 Fig).

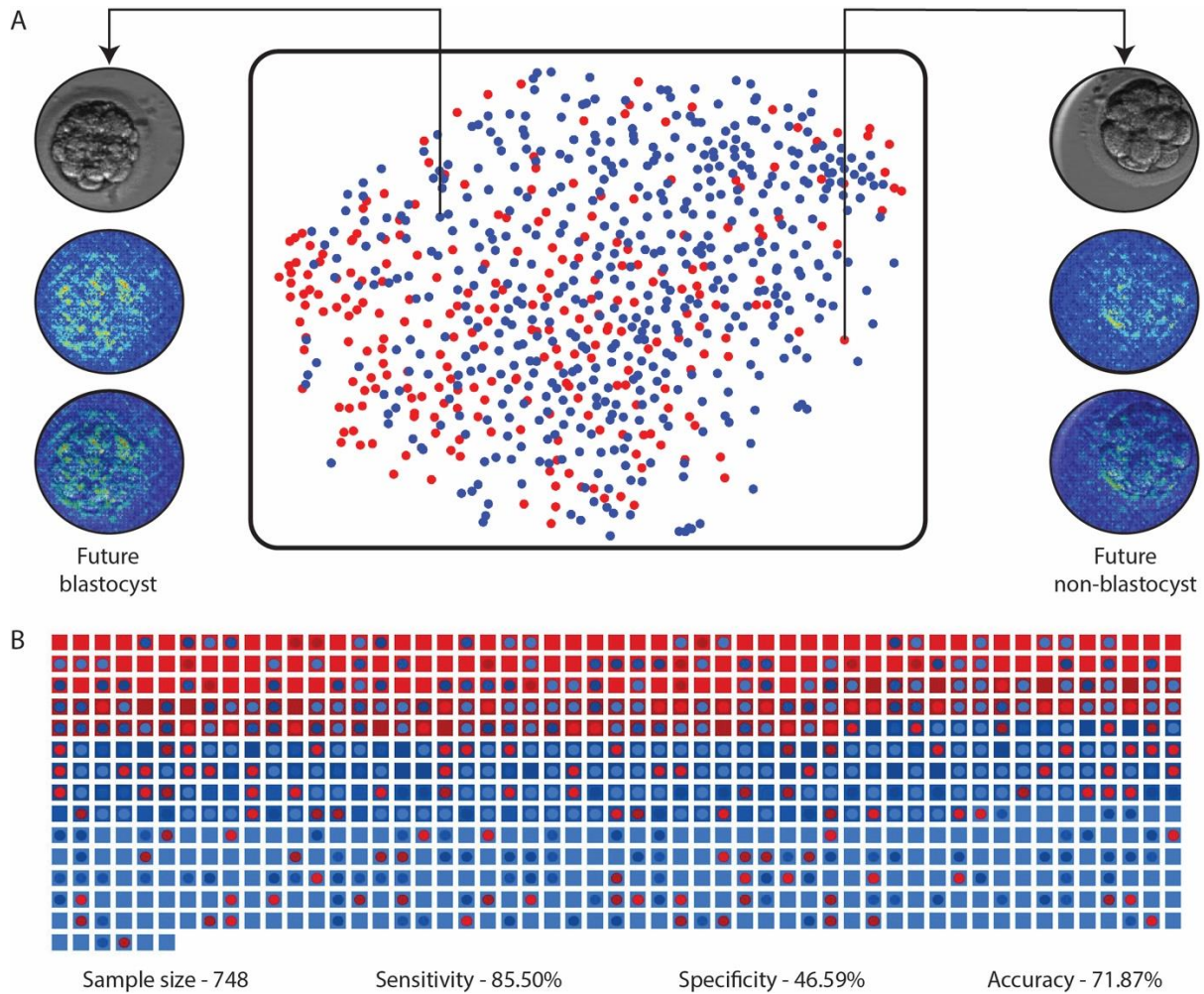

**Figure 3. Evaluation at the cleavage stage.** (A) The t-SNE plot for the Xception model trained to classify between embryos that do not develop to blastocyst stage (classes 1 and 2) and those that develop to the blastocyst stage (classes 3, 4, and 5). The saliency map of the two groups provides an example of the features that the network uses to classify embryos at 70 hpi. (B) The dot matrix plot illustrates the performance of the AI system in evaluating embryos (n=748) from the test set of 97 patients. The squares represent true labels and the circles within them represent the system's classification. Blue squares and circles represent embryos that developed to the blastocyst stage by 113 hpi while red squares and circles represent embryos that do not develop into blastocysts by 113 hpi. The color gradient differentiates the subclasses of the two major groups.

The sensitivity and specificity of the algorithm for blastocyst prediction (between the 2 inference classes) at 70 hpi were 85.50% (CI: 81.95% to 88.58%) and 46.49% (CI: 40.26% to 52.99%), respectively when using the test set of 748. The PPV and NPV for blastocyst prediction at 70 hpi were 74.81% (CI: 72.44% to 77.04%) and 63.39% (CI: 57.21% to 69.15%), respectively (Fig 3B, Table S3). The AUC value established through a ROC analysis was 0.685 (CI: 0.650 to 0.718) (S3B Fig).

## 3.2 System evaluation for embryo selection and comparisons with embryologists

To evaluate the potential improvement in raw predictive power, we compared the accuracy of predictions by embryologists, collected retrospectively during routine clinical analysis, in identifying embryos that will eventually develop into blastocysts when presented with embryo morphology imaged at 44 hpi and 70 hpi (n=803). We also evaluated their performance with and without time lapse specific information P2 (duration of the 2-cell stage) and P3 (duration of the three-cell stage) (n=803) and compared their overall prediction performance against the performance of the trained neural network (n=748) (S5 Fig, Table S2). The neural network significantly outperformed a typical embryologist's performance, in terms identifying embryos that develop into blastocysts correctly ($P < 0.0001$; N-1 Chi-squared test) and the overall accuracy in prediction ($P < 0.0001$; N-1 Chi-squared test), regardless of the evaluated methodology. A similar skew in prediction was observed in amongst embryologists. However, the skew indicated that methods used by embryologists identified non-blastocysts better than blastocysts in our evaluations.

According to the American Society for Reproductive Medicine guidelines on the limits to the number of embryo transfer, 1 embryo is transferred for high prognosis patients with <37 years of age and 2 or more embryos are transferred for patients with >37 years of age as well as younger patients with low prognosis (Guidance on the limits to the number of embryos to transfer: a committee opinion, 2017). All currently available systems for embryo imaging do not offer top-quality embryo selections and thus embryologists are required to devote time into evaluating each embryo's predicted potential prior to selection. Selections become challenging especially at the cleavage stage of development due to the limited set of embryo features that can be used to gauge its developmental potential. The developed CNN calculates the probability of each embryo belonging to one of the 5 training classes for any given patient embryo cohort. However, clinically, a continuous metric-based scoring system is more ideal for ranking embryos towards establishing the order of transfer. In order to use the five probability values effectively for calculating the embryo score, we utilized a genetic algorithm, which is well-suited for optimization problems with multiple existing solutions. Here, the genetic algorithm (GA) empowered the developed CNN to make selections of the top-quality embryos within a patient's embryo cohort at 70 hpi. The criteria for selecting top-quality embryos using the CNN+GA system and manual assessment are discussed in the methods section. Using a test set of 97 patient embryo cohorts in a double-blinded study; we compared the accuracy of the CNN+GA selections with the accuracy of the manual selections performed by five embryologists for the scenarios of single embryo transfers (SET) and double embryo transfers (DET). We evaluated the ability of such a combination of algorithms to effectively select (i) blastocyst(s) for transfer and (ii) the highest quality of blastocyst(s) available for transfer.

The accuracy of the CNN+GA in selecting an embryo, which developed into a blastocyst by 113 hpi for SET at 70 hpi was 83.51% while, the average accuracy of the embryologists (n=5) was 80.21% (CI: 77.74% to 82.67%) (Fig 4A, Table 1). A one sample t-test revealed that the system performed significantly ($P<0.05$) better than the embryologists in selecting a single cleavage stage embryo for transfer that will eventually develop into a blastocyst. When the CNN+GA and the embryologists were allowed to select two embryos of which at least one developed into a blastocyst by 113 hpi for a DET at 70 hpi, the CNN+GA performed with an accuracy of 96.90% while the embryologists performed with an average accuracy of 94.02% (CI: 92.35% to 95.69%) (Fig 4B,

Table 1). A one sample t-test revealed that the system performed significantly (P<0.05) better than embryologists in selecting two embryos for transfer among which at least one would eventually form a blastocyst. The accuracy of the CNN+GA in selecting an embryo at 70 hpi, which developed into a high-quality blastocyst (HQB) for a SET, was 63.91% that is significantly higher (P<0.05) than the average accuracy of the embryologists (52.78%, CI: 48.60% to 56.97%) (Fig 4C, Table 1). The accuracy of the CNN+GA in selecting an embryo at 70 hpi, which developed into HQB for a DET, was significantly higher (79.38%, P<0.05) compared to the embryologists with an average accuracy of 72.37% (CI: 70.70% to 74.04%) (Fig 4D, Table 1).

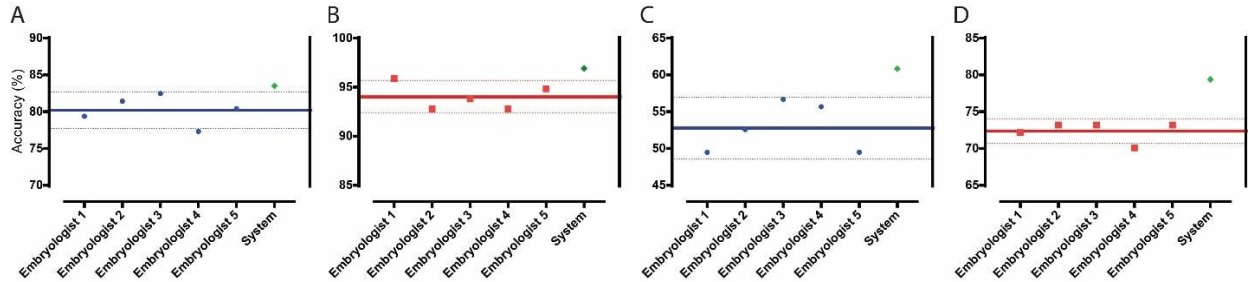

**Fig 4. Comparisons of performance between embryologists and the neural network scheme in selecting embryos.** (A) shows the performance of embryologists and the algorithm in selecting a single embryo (SET) at 70 hpi that eventually developed into a blastocyst. (B) shows their performance when two selections (DET) were allowed at 70 hpi and where at least one transformed into a blastocyst. (C) shows the performance of embryologists and the algorithm in selecting a single embryo (SET) at 70 hpi that eventually developed into a high-quality blastocyst. (D) shows their performance when two selections (DET) were allowed at 70 hpi and where at least one transformed into a high-quality blastocyst.

|  | Day 3 (70 hpi) HQB selection | | Day 3 (70 hpi) (Blastocyst) | |
|---|---|---|---|---|
|  | SET | DET | SET | DET |
| AI system accuracy (%) | 63.91 | 79.38 | 83.51 | 96.90 |
| Embryologist average accuracy (%) | 52.78 | 72.37 | 80.21 | 94.02 |

**Table 1 The performance of the embryologists and the neural network in embryo selection.** The embryo selections by the embryologists and the network were made using a dataset of 97 patient embryos. HQB stands for highest quality blastocyst, which refers to embryos of class 5 (>3cc blastocyst quality).

## 4. Discussion

The developed neural network automatically and without any human intervention analyzes embryo images recorded at a single time point (70hpi) to predict embryo developmental fate. Current time lapse systems for embryo assessment make use of data collected over multiple days during embryo in vitro culture to study their morphological and morphokinetic features. The work reported here showcases the feasibility of using embryo morphological data on a single 2D embryo image for embryo assessment and its developmental fate prediction.

Currently, the main criterion in selecting an embryo at the cleavage stage for transfer to a patient is mainly based on morphology assessment at 70 hpi without considering the embryo's potential in developing to blastocysts. The saliency maps revealed localization of the highest weighted system-selected features around specific regions of the embryo morphology used by embryologists in assessing embryos. The 70 hpi saliency maps, revealed scattered highlights over the embryos, making it difficult to ascertain all the specific regions of interest utilized by the network (S2 Fig). Blastomeres and regions of cellular fragmentation were highlighted in most embryos. Such technology advancement sets a milestone for artificial intelligence (AI) in embryology since currently the embryo features used by embryologists and semi-automated systems for embryo assessment are selected through manual observations alone.

TLI systems in general have a high specificity and critically low sensitivity in predicting blastocyst development (Petersen et al., 2016; Kirkegaard et al., 2015). These systems can identify embryos that arrest during embryo development but are unable to predict those that develop to the blastocyst stage. This is a significant drawback since successful embryo transfers require identification of the highest quality viable embryo. In contrast, the reported AI technology can better identify most embryos that eventually become blastocysts by 113 hpi. The network performs well in identifying classes 1 and 5. The confusion matrix shows a bias between these classes with a stronger bias towards class 5 (S4 Fig). High sensitivity in identifying blastocysts may be likely due to 70 hpi embryos not presenting features indicative of stunted developmental growth leading their eventual development into embryos belonging to classes 2, 3, and 4 by 113 hpi, and can lead to more embryos being predicted to be class 5. On the other hand, the low specificity during cleavage stage blastocyst prediction likely stems from the confusion that exists between classes 2 and 3 based on the dispersed feature map, t-SNE clusters, and confusion matrices (S4 Fig and Fig 3). Although the system has poor specificity during absolute scoring, the high sensitivity implies that most viable embryos are utilized in the selection process. Since clinically only the highest quality embryo(s) (HQB) is transferred to the patient, the genetic algorithm filters lower confidence embryos and selects eventual HQBs with higher accuracy compared to embryologists. The genetic algorithm helps to overcome scoring errors even when evaluating embryos at 113 hpi. The highest frequency of errors occurs between classes 2 and 3, due to using the genetic algorithm, did not seem to hold significant effect on the selection outcomes.

All previously developed computer-assisted approaches for embryo assessment demand expensive and specialized optical systems. The reported CNN-based approach makes use of single timepoint static images and therefore is adaptable to data collected using traditional instruments. Therefore, the CNN can be easily augmented into current clinical practice equipped with basic desktop bright-field digital microscopes.

Advances in AI methodologies have paved the way for a new direction of computer-assisted decision support systems. Our AI-based approach requires no handcrafted feature selection, and the algorithm could learn and identify important embryo features for automated assessment. The reported AI-based approach can evaluate embryos at 70 hpi with consistent performance significantly better than typical conventional assessments by embryologists. Furthermore, the CNN was able to outperform embryologists in the selection of the highest quality embryos. Currently, since the network is only using a single timepoint image in predictions, its predictive power can be vastly increased by augmenting patient specific data along with time-lapse data during its training. Training such an algorithm would need significantly more data than the 2446 patient embryos that were used for this study. Our work, however, sets a milestone in cleavage stage predictions and warrants further research in improving the predictive power using much larger datasets. While blastocyst development is an important outcome for prediction in IVF for cleavage stage transfers, performance in the prediction of pregnancy and live birth is a more lucrative metric to train against, a limitation in this study. Our future direction of research will be directed towards more developing neural networks more personalized to patients in predicting pregnancy or implantation.

## Acknowledgements

The authors would like to thank embryology staff from Massachusetts General Hospital for participating in this study. This work was supported by the Brigham and Women's Hospital (BWH) Precision Medicine Developmental Award (BWH Precision Medicine Program) and Partners Innovation Discovery Grant (Partners Healthcare). It was also partially supported through 1R01AI118502, and R01AI138800 Awards (National Institute of Health).

**Author contributions**: Manoj Kumar Kanakasabapathy, Charles L Bormann, Prudhvi Thirumalaraju and Hadi Shafiee designed the study. Hadi Shafiee supervised the study. Charles L Bormann, Irene Souter and Irene Dimitriadis supervised and assisted data annotations and data collection at the fertility center. Prudhvi Thirumalaraju, Raghav Gupta and Rohan Pooniwala. trained and developed the neural networks. Raghav Gupta and Hemanth Kandula developed the smartphone applications. Manoj Kumar Kanakasabapathy, Prudhvi Thirumalaraju and Charles L Bormann analyzed the data. Manoj Kumar Kanakasabapathy and Hadi Shafiee wrote the manuscript. All co-authors edited the manuscript.

**Data and materials availability**: Restrictions apply to the availability of the medical training/validation data, which were used with permission for the current study, and so are not publicly available. Some data may be available from the authors upon reasonable request and with permission of the Massachusetts General Hospital.

# Supplementary information

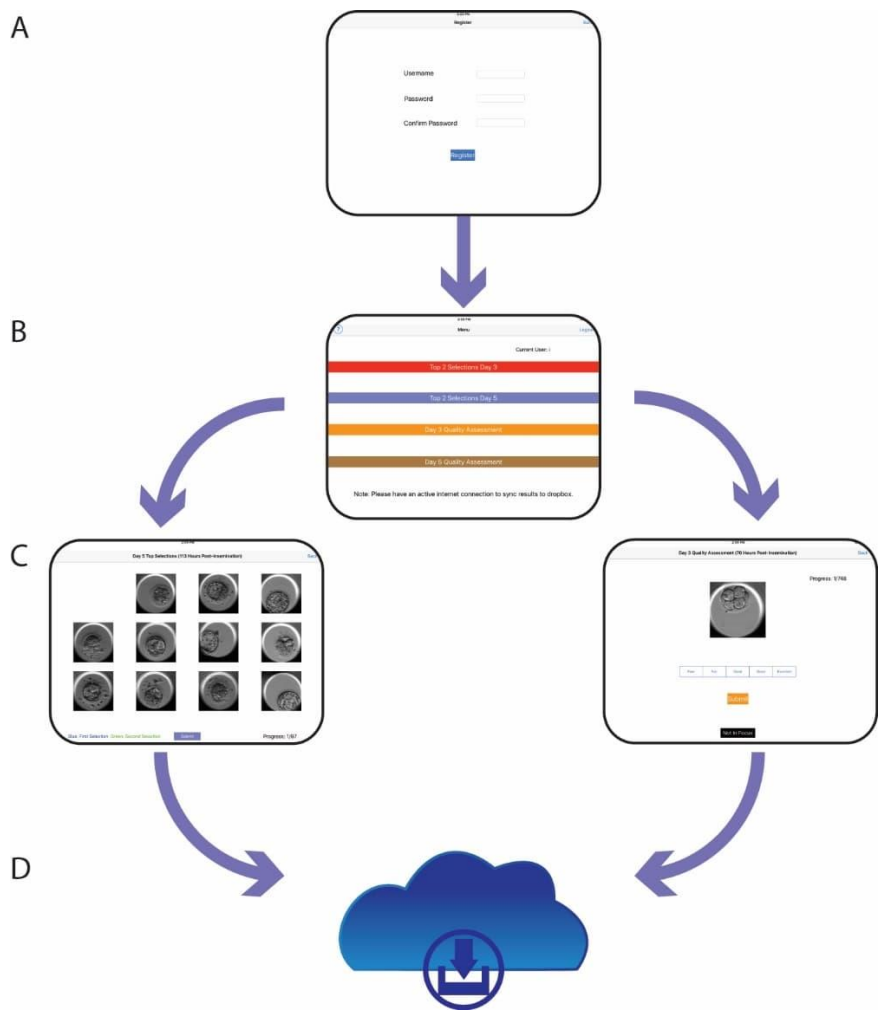

**S1 Fig. Software application schematic.** The schematic illustrates the dataflow from embryologist to our repository when performing selections and classifications. (A) The login screen which provides access only to the user performing the annotations and minimizes data mix-ups. (B) The main menu screen where the user was able to select the task of selection or classification. (C) and (D) The classification and selection activities, when the user selects their annotation, the system records and sends a copy to the cloud asynchronously.

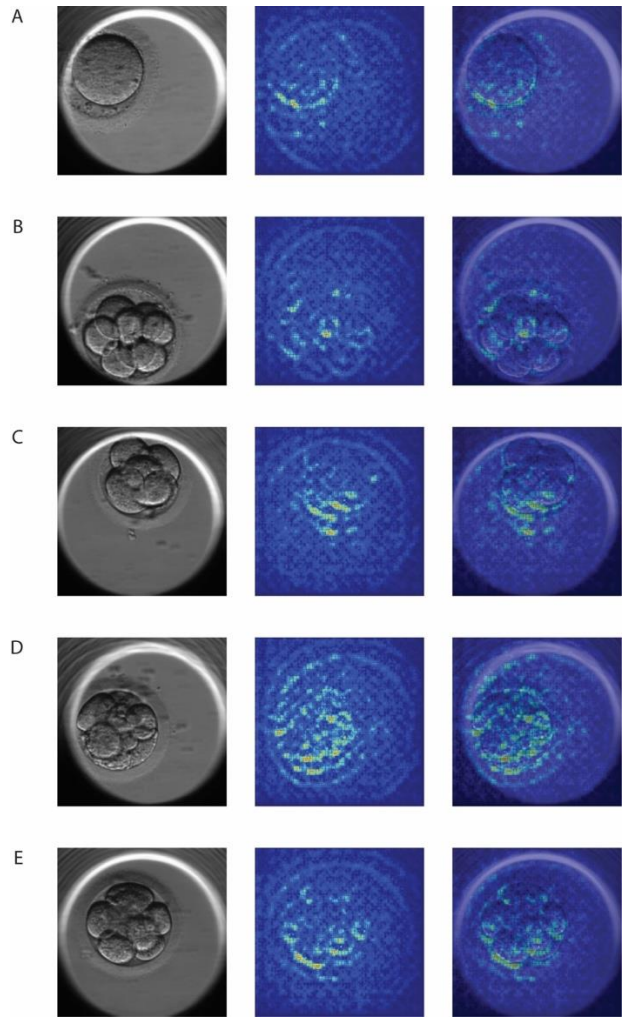

**S2 Fig. Saliency maps of embryos assessed at 70 hours post-insemination (hpi).** The saliency map was extracted from the network to highlight the highest weighted features for the image. (A) An embryo that eventually turned into class 1 embryo category at 113 hpi along with its respective saliency map and saliency map overlaid on the bright-field embryoscope image (B) An embryo that eventually turned into class 2 embryo category at 113 hpi along with its respective saliency map and saliency map overlaid on the bright-field embryoscope image. (C) An embryo that eventually turned into class 3 embryo category at 113 hpi along with its respective saliency map and saliency map overlaid on the bright-field embryoscope image. (D) An embryo that eventually turned into class 4 embryo category at 113 hpi along with its respective saliency map and saliency map overlaid on the bright-field embryoscope image. (E) An embryo that eventually turned into class 5 embryo category at 113 hpi along with its respective saliency map and saliency map overlaid on the bright-field embryoscope image. Blue indicates the lowest weights while red indicates the highest weights. Blastomeres and regions of cellular fragmentation were highlighted in most embryos.

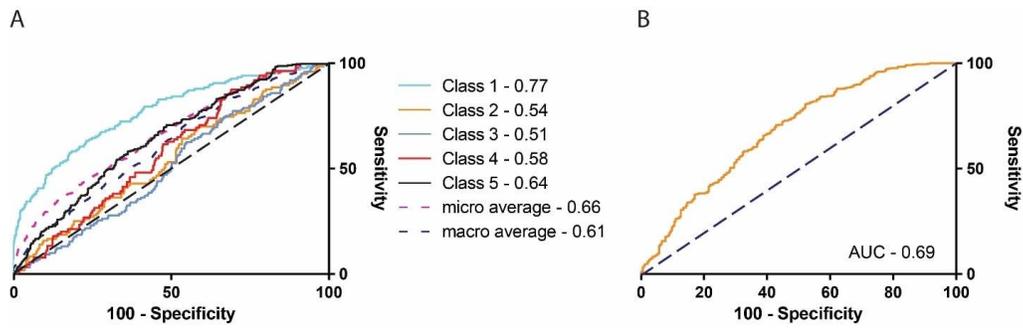

**S3 Fig. ROC analysis for classification and prediction tasks.** (A) ROC curves for 5 class at 70 hpi. (B) ROC curves for 2 classifications at 70 hpi.

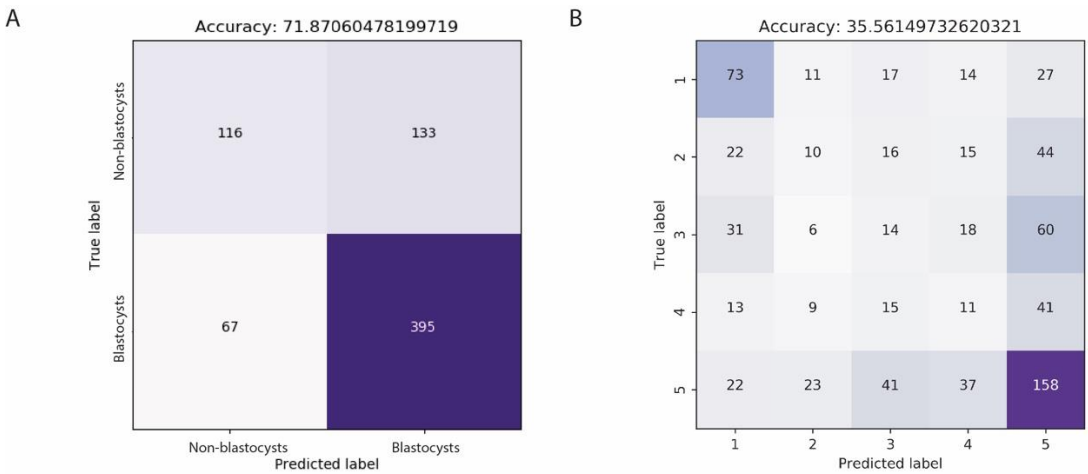

**S4 Fig. Confusion Matrices of the neural network in embryo classification tasks.** (A) Confusion matrix for predicting embryos between 2 classes using 70 hpi embryo images. (B) Confusion matrix for predicting embryos between 5 classes using 70 hpi embryo images.

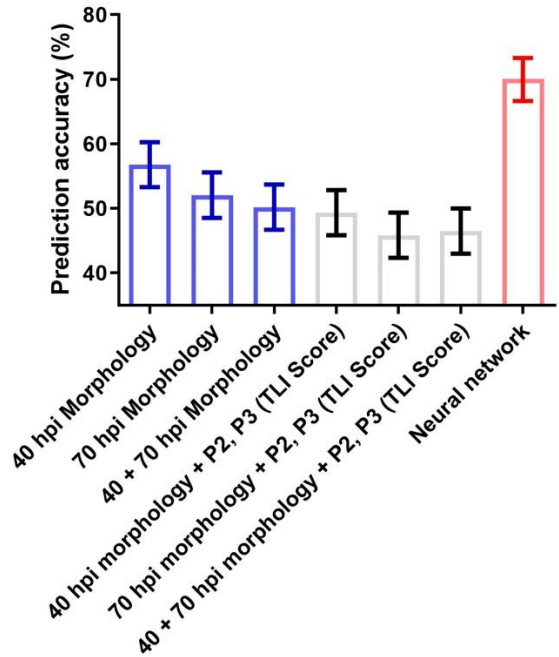

**S5 Fig. Comparison of overall predictive performance between embryologists and the neural network.**

The embryologist performance data in predicting blastocyst development using pre-blastocyst images reflects the clinical performance across 8 embryologists and was collected retrospectively during routine clinical analyses (n=803). The neural network accuracy is the performance of the neural network using the test set (n=748). The error bars represent 95% confidence intervals (Clopper-Pearson).

| Dataset | Total number |
|---|---|
| Embryos | 803 |
| Blastocysts | 527 |
| Non-blastocysts | 276 |

**Table S1: Dataset Information**

| Method of Cleavage Stage Top Embryo Quality Selection | Total Embryos predicted to be of the highest quality | Total selected embryos that developed into blastocysts | Total selected embryos that developed into non-blastocysts | Total unselected embryos that developed into blastocysts | Total unselected embryos that developed into non-blastocysts |
|---|---|---|---|---|---|
| 44 hpi images alone | 254 | 217 | 37 | 310 | 239 |
| 70 hpi images alone | 192 | 167 | 25 | 360 | 251 |
| 44 + 70 hpi images | 153 | 140 | 13 | 387 | 263 |
| 44 hpi images + P2, P3 (TLI Score) | 172 | 146 | 26 | 381 | 250 |
| 70 hpi images + P2, P3 (TLI Score) | 132 | 112 | 20 | 415 | 256 |
| 44 + 70 hpi images + P2, P3 (TLI Score) | 121 | 109 | 12 | 418 | 264 |

**Table S2: Raw data of embryologist's predictions collected during clinical analysis of embryos.** The data was collected during routine clinical evaluations using 44 and 70hpi images, and time-lapse information such as P2 (duration of the 2-cell stage) and P3 (duration of the three-cell stage).

| 70 hpi – System Predictions | | | | | | | | |
|---|---|---|---|---|---|---|---|---|
| **Embryo #** | **Real class** | **System Prediction** | **Embryo #** | **Real class** | **System Prediction** | **Embryo #** | **Real class** | **System Prediction** |
| Embryo 1 | 3 | 5 | Embryo 2 | 4 | 5 | Embryo 3 | 5 | 5 |
| Embryo 4 | 2 | 3 | Embryo 5 | 4 | 2 | Embryo 6 | 5 | 3 |
| Embryo 7 | 4 | 1 | Embryo 8 | 5 | 5 | Embryo 9 | 5 | 4 |
| Embryo 10 | 5 | 5 | Embryo 11 | 1 | 1 | Embryo 12 | 1 | 5 |
| Embryo 13 | 5 | 5 | Embryo 14 | 5 | 2 | Embryo 15 | 2 | 1 |
| Embryo 16 | 4 | 1 | Embryo 17 | 5 | 5 | Embryo 18 | 5 | 4 |
| Embryo 19 | 5 | 5 | Embryo 20 | 3 | 1 | Embryo 21 | 2 | 5 |
| Embryo 22 | 5 | 5 | Embryo 23 | 3 | 5 | Embryo 24 | 4 | 5 |
| Embryo 25 | 5 | 1 | Embryo 26 | 4 | 3 | Embryo 27 | 3 | 3 |
| Embryo 28 | 5 | 4 | Embryo 29 | 4 | 3 | Embryo 30 | 5 | 3 |
| Embryo 31 | 5 | 5 | Embryo 32 | 5 | 3 | Embryo 33 | 5 | 4 |
| Embryo 34 | 2 | 4 | Embryo 35 | 4 | 5 | Embryo 36 | 4 | 3 |
| Embryo 37 | 4 | 5 | Embryo 38 | 3 | 4 | Embryo 39 | 5 | 5 |
| Embryo 40 | 5 | 5 | Embryo 41 | 5 | 5 | Embryo 42 | 5 | 2 |
| Embryo 43 | 5 | 5 | Embryo 44 | 5 | 5 | Embryo 45 | 3 | 3 |
| Embryo 46 | 5 | 5 | Embryo 47 | 5 | 5 | Embryo 48 | 4 | 5 |
| Embryo 49 | 2 | 3 | Embryo 50 | 4 | 1 | Embryo 51 | 3 | 4 |
| Embryo 52 | 1 | 3 | Embryo 53 | 1 | 1 | Embryo 54 | 1 | 1 |
| Embryo 55 | 1 | 3 | Embryo 56 | 1 | 4 | Embryo 57 | 2 | 4 |
| Embryo 58 | 1 | 4 | Embryo 59 | 2 | 2 | Embryo 60 | 1 | 5 |
| Embryo 61 | 2 | 2 | Embryo 62 | 5 | 5 | Embryo 63 | 1 | 1 |
| Embryo 64 | 2 | 5 | Embryo 65 | 2 | 5 | Embryo 66 | 1 | 5 |
| Embryo 67 | 1 | 1 | Embryo 68 | 1 | 1 | Embryo 69 | 1 | 2 |
| Embryo 70 | 3 | 1 | Embryo 71 | 4 | 2 | Embryo 72 | 4 | 4 |
| Embryo 73 | 3 | 3 | Embryo 74 | 5 | 5 | Embryo 75 | 5 | 5 |
| Embryo 76 | 3 | 1 | Embryo 77 | 5 | 3 | Embryo 78 | 5 | 5 |
| Embryo 79 | 5 | 2 | Embryo 80 | 1 | 2 | Embryo 81 | 4 | 3 |
| Embryo 82 | 1 | 5 | Embryo 83 | 5 | 5 | Embryo 84 | 3 | 4 |
| Embryo 85 | 5 | 5 | Embryo 86 | 1 | 4 | Embryo 87 | 3 | 1 |
| Embryo 88 | 4 | 5 | Embryo 89 | 4 | 4 | Embryo 90 | 5 | 5 |
| Embryo 91 | 3 | 1 | Embryo 92 | 2 | 5 | Embryo 93 | 5 | 5 |
| Embryo 94 | 4 | 1 | Embryo 95 | 3 | 2 | Embryo 96 | 3 | 1 |
| Embryo 97 | 4 | 5 | Embryo 98 | 1 | 1 | Embryo 99 | 4 | 5 |
| Embryo 100 | 5 | 2 | Embryo 101 | 1 | 1 | Embryo 102 | 3 | 1 |
| Embryo 103 | 4 | 2 | Embryo 104 | 3 | 1 | Embryo 105 | 2 | 2 |
| Embryo 106 | 1 | 1 | Embryo 107 | 1 | 1 | Embryo 108 | 3 | 3 |
| Embryo 109 | 2 | 5 | Embryo 110 | 1 | 1 | Embryo 111 | 3 | 5 |
| Embryo 112 | 1 | 1 | Embryo 113 | 5 | 5 | Embryo 114 | 2 | 5 |
| Embryo 115 | 2 | 5 | Embryo 116 | 1 | 1 | Embryo 117 | 5 | 3 |
| Embryo 118 | 1 | 4 | Embryo 119 | 5 | 1 | Embryo 120 | 4 | 2 |
| Embryo 121 | 2 | 5 | Embryo 122 | 3 | 4 | Embryo 123 | 5 | 4 |
| Embryo 124 | 5 | 5 | Embryo 125 | 5 | 5 | Embryo 126 | 2 | 4 |
| Embryo 127 | 5 | 5 | Embryo 128 | 3 | 3 | Embryo 129 | 5 | 5 |
| Embryo 130 | 2 | 2 | Embryo 131 | 4 | 5 | Embryo 132 | 4 | 5 |
| Embryo 133 | 5 | 1 | Embryo 134 | 3 | 5 | Embryo 135 | 3 | 5 |
| Embryo 136 | 4 | 5 | Embryo 137 | 3 | 5 | Embryo 138 | 4 | 4 |
| Embryo 139 | 1 | 1 | Embryo 140 | 3 | 4 | Embryo 141 | 5 | 2 |
| Embryo 142 | 2 | 4 | Embryo 143 | 5 | 5 | Embryo 144 | 1 | 1 |
| Embryo 145 | 5 | 1 | Embryo 146 | 3 | 5 | Embryo 147 | 2 | 4 |
| Embryo 148 | 1 | 5 | Embryo 149 | 5 | 5 | Embryo 150 | 1 | 5 |
| Embryo 151 | 1 | 4 | Embryo 152 | 1 | 1 | Embryo 153 | 3 | 1 |
| Embryo 154 | 5 | 1 | Embryo 155 | 1 | 1 | Embryo 156 | 1 | 1 |
| Embryo 157 | 4 | 2 | Embryo 158 | 4 | 4 | Embryo 159 | 5 | 3 |
| Embryo 160 | 1 | 4 | Embryo 161 | 1 | 1 | Embryo 162 | 1 | 1 |
| Embryo 163 | 5 | 5 | Embryo 164 | 5 | 4 | Embryo 165 | 5 | 3 |
| Embryo 166 | 4 | 3 | Embryo 167 | 1 | 1 | Embryo 168 | 5 | 1 |
| Embryo 169 | 3 | 1 | Embryo 170 | 4 | 1 | Embryo 171 | 2 | 5 |
| Embryo 172 | 4 | 2 | Embryo 173 | 4 | 4 | Embryo 174 | 5 | 5 |
| Embryo 175 | 4 | 3 | Embryo 176 | 5 | 3 | Embryo 177 | 5 | 3 |
| Embryo 178 | 5 | 5 | Embryo 179 | 4 | 5 | Embryo 180 | 5 | 5 |
| Embryo 181 | 5 | 4 | Embryo 182 | 4 | 1 | Embryo 183 | 4 | 3 |
| Embryo 184 | 4 | 4 | Embryo 185 | 3 | 2 | Embryo 186 | 5 | 3 |
| Embryo 187 | 4 | 3 | Embryo 188 | 5 | 1 | Embryo 189 | 4 | 5 |
| Embryo 190 | 5 | 5 | Embryo 191 | 4 | 1 | Embryo 192 | 3 | 3 |
| Embryo 193 | 5 | 5 | Embryo 194 | 5 | 5 | Embryo 195 | 2 | 3 |
| Embryo 196 | 5 | 3 | Embryo 197 | 1 | 1 | Embryo 198 | 4 | 5 |
| Embryo 199 | 5 | 5 | Embryo 200 | 4 | 3 | Embryo 201 | 2 | 5 |
| Embryo 202 | 3 | 1 | Embryo 203 | 5 | 5 | Embryo 204 | 1 | 4 |
| Embryo 205 | 5 | 2 | Embryo 206 | 1 | 5 | Embryo 207 | 1 | 1 |
| Embryo 208 | 1 | 4 | Embryo 209 | 1 | 5 | Embryo 210 | 3 | 1 |
| Embryo 211 | 5 | 5 | Embryo 212 | 5 | 5 | Embryo 213 | 3 | 1 |
| Embryo 214 | 3 | 5 | Embryo 215 | 5 | 5 | Embryo 216 | 3 | 5 |
| Embryo 217 | 5 | 4 | Embryo 218 | 4 | 2 | Embryo 219 | 5 | 3 |
| Embryo 220 | 5 | 5 | Embryo 221 | 5 | 1 | Embryo 222 | 3 | 1 |

| Embryo 223 | 3 | 5 | Embryo 224 | 3 | 5 | Embryo 225 | 4 | 5 |
|---|---|---|---|---|---|---|---|---|
| Embryo 226 | 2 | 5 | Embryo 227 | 5 | 5 | Embryo 228 | 5 | 5 |
| Embryo 229 | 1 | 4 | Embryo 230 | 2 | 5 | Embryo 231 | 1 | 3 |
| Embryo 232 | 3 | 5 | Embryo 233 | 5 | 5 | Embryo 234 | 3 | 5 |
| Embryo 235 | 3 | 5 | Embryo 236 | 2 | 1 | Embryo 237 | 3 | 5 |
| Embryo 238 | 2 | 5 | Embryo 239 | 5 | 5 | Embryo 240 | 2 | 1 |
| Embryo 241 | 3 | 5 | Embryo 242 | 3 | 5 | Embryo 243 | 1 | 1 |
| Embryo 244 | 2 | 2 | Embryo 245 | 2 | 1 | Embryo 246 | 3 | 1 |
| Embryo 247 | 1 | 1 | Embryo 248 | 4 | 5 | Embryo 249 | 2 | 1 |
| Embryo 250 | 5 | 3 | Embryo 251 | 5 | 5 | Embryo 252 | 5 | 2 |
| Embryo 253 | 5 | 5 | Embryo 254 | 2 | 4 | Embryo 255 | 3 | 5 |
| Embryo 256 | 4 | 5 | Embryo 257 | 1 | 3 | Embryo 258 | 4 | 5 |
| Embryo 259 | 4 | 5 | Embryo 260 | 4 | 5 | Embryo 261 | 5 | 5 |
| Embryo 262 | 3 | 5 | Embryo 263 | 2 | 5 | Embryo 264 | 3 | 2 |
| Embryo 265 | 3 | 5 | Embryo 266 | 2 | 2 | Embryo 267 | 5 | 5 |
| Embryo 268 | 2 | 1 | Embryo 269 | 5 | 3 | Embryo 270 | 3 | 5 |
| Embryo 271 | 3 | 5 | Embryo 272 | 1 | 5 | Embryo 273 | 5 | 4 |
| Embryo 274 | 1 | 4 | Embryo 275 | 2 | 4 | Embryo 276 | 3 | 3 |
| Embryo 277 | 3 | 5 | Embryo 278 | 1 | 1 | Embryo 279 | 1 | 5 |
| Embryo 280 | 2 | 3 | Embryo 281 | 2 | 3 | Embryo 282 | 2 | 5 |
| Embryo 283 | 5 | 4 | Embryo 284 | 3 | 1 | Embryo 285 | 2 | 1 |
| Embryo 286 | 1 | 1 | Embryo 287 | 1 | 1 | Embryo 288 | 1 | 1 |
| Embryo 289 | 1 | 1 | Embryo 290 | 3 | 2 | Embryo 291 | 5 | 3 |

| Embryo 292 | 1 | 5 | Embryo 293 | 2 | 5 | Embryo 294 | 1 | 3 |
|---|---|---|---|---|---|---|---|---|
| Embryo 295 | 1 | 5 | Embryo 296 | 2 | 5 | Embryo 297 | 3 | 5 |
| Embryo 298 | 2 | 1 | Embryo 299 | 5 | 5 | Embryo 300 | 1 | 2 |
| Embryo 301 | 5 | 4 | Embryo 302 | 4 | 3 | Embryo 303 | 3 | 2 |
| Embryo 304 | 2 | 5 | Embryo 305 | 2 | 3 | Embryo 306 | 1 | 1 |
| Embryo 307 | 5 | 5 | Embryo 308 | 4 | 5 | Embryo 309 | 3 | 3 |
| Embryo 310 | 3 | 5 | Embryo 311 | 4 | 5 | Embryo 312 | 5 | 5 |
| Embryo 313 | 4 | 1 | Embryo 314 | 2 | 1 | Embryo 315 | 1 | 1 |
| Embryo 316 | 2 | 1 | Embryo 317 | 1 | 5 | Embryo 318 | 1 | 1 |
| Embryo 319 | 5 | 5 | Embryo 320 | 1 | 4 | Embryo 321 | 5 | 5 |
| Embryo 322 | 2 | 5 | Embryo 323 | 5 | 5 | Embryo 324 | 5 | 5 |
| Embryo 325 | 5 | 4 | Embryo 326 | 3 | 1 | Embryo 327 | 1 | 3 |
| Embryo 328 | 4 | 5 | Embryo 329 | 3 | 5 | Embryo 330 | 3 | 5 |
| Embryo 331 | 4 | 5 | Embryo 332 | 5 | 5 | Embryo 333 | 2 | 5 |
| Embryo 334 | 5 | 5 | Embryo 335 | 3 | 5 | Embryo 336 | 3 | 5 |
| Embryo 337 | 1 | 3 | Embryo 338 | 2 | 1 | Embryo 339 | 2 | 5 |
| Embryo 340 | 5 | 4 | Embryo 341 | 3 | 3 | Embryo 342 | 4 | 5 |
| Embryo 343 | 1 | 3 | Embryo 344 | 2 | 3 | Embryo 345 | 4 | 5 |
| Embryo 346 | 1 | 2 | Embryo 347 | 5 | 5 | Embryo 348 | 4 | 2 |
| Embryo 349 | 4 | 5 | Embryo 350 | 5 | 5 | Embryo 351 | 5 | 4 |
| Embryo 352 | 5 | 5 | Embryo 353 | 2 | 1 | Embryo 354 | 1 | 5 |
| Embryo 355 | 5 | 5 | Embryo 356 | 1 | 1 | Embryo 357 | 5 | 4 |
| Embryo 358 | 1 | 1 | Embryo 359 | 2 | 4 | Embryo 360 | 3 | 3 |
| Embryo 361 | 1 | 5 | Embryo 362 | 5 | 1 | Embryo 363 | 5 | 5 |
| Embryo 364 | 4 | 4 | Embryo 365 | 1 | 1 | Embryo 366 | 5 | 5 |
| Embryo 367 | 1 | 5 | Embryo 368 | 2 | 5 | Embryo 369 | 1 | 1 |
| Embryo 370 | 1 | 1 | Embryo 371 | 5 | 4 | Embryo 372 | 5 | 5 |
| Embryo 373 | 5 | 3 | Embryo 374 | 5 | 4 | Embryo 375 | 4 | 5 |
| Embryo 376 | 5 | 5 | Embryo 377 | 5 | 5 | Embryo 378 | 5 | 5 |
| Embryo 379 | 3 | 1 | Embryo 380 | 5 | 4 | Embryo 381 | 5 | 5 |
| Embryo 382 | 5 | 3 | Embryo 383 | 3 | 4 | Embryo 384 | 5 | 5 |
| Embryo 385 | 3 | 5 | Embryo 386 | 5 | 2 | Embryo 387 | 5 | 5 |
| Embryo 388 | 2 | 5 | Embryo 389 | 5 | 5 | Embryo 390 | 4 | 3 |
| Embryo 391 | 5 | 5 | Embryo 392 | 5 | 5 | Embryo 393 | 5 | 5 |
| Embryo 394 | 3 | 1 | Embryo 395 | 2 | 1 | Embryo 396 | 5 | 1 |
| Embryo 397 | 1 | 1 | Embryo 398 | 5 | 5 | Embryo 399 | 5 | 5 |
| Embryo 400 | 5 | 5 | Embryo 401 | 5 | 4 | Embryo 402 | 5 | 4 |
| Embryo 403 | 5 | 2 | Embryo 404 | 4 | 5 | Embryo 405 | 5 | 5 |
| Embryo 406 | 5 | 5 | Embryo 407 | 1 | 1 | Embryo 408 | 5 | 4 |
| Embryo 409 | 5 | 5 | Embryo 410 | 5 | 5 | Embryo 411 | 5 | 4 |
| Embryo 412 | 4 | 5 | Embryo 413 | 5 | 5 | Embryo 414 | 1 | 3 |
| Embryo 415 | 5 | 5 | Embryo 416 | 2 | 3 | Embryo 417 | 1 | 1 |
| Embryo 418 | 3 | 3 | Embryo 419 | 2 | 1 | Embryo 420 | 5 | 5 |
| Embryo 421 | 5 | 5 | Embryo 422 | 3 | 5 | Embryo 423 | 4 | 4 |
| Embryo 424 | 1 | 3 | Embryo 425 | 1 | 1 | Embryo 426 | 3 | 5 |
| Embryo 427 | 2 | 5 | Embryo 428 | 1 | 1 | Embryo 429 | 4 | 5 |
| Embryo 430 | 1 | 1 | Embryo 431 | 3 | 5 | Embryo 432 | 5 | 5 |
| Embryo 433 | 5 | 5 | Embryo 434 | 3 | 4 | Embryo 435 | 2 | 5 |
| Embryo 436 | 3 | 5 | Embryo 437 | 5 | 4 | Embryo 438 | 3 | 3 |
| Embryo 439 | 1 | 2 | Embryo 440 | 3 | 5 | Embryo 441 | 2 | 4 |
| Embryo 442 | 3 | 4 | Embryo 443 | 2 | 1 | Embryo 444 | 2 | 1 |
| Embryo 445 | 5 | 5 | Embryo 446 | 1 | 1 | Embryo 447 | 2 | 2 |
| Embryo 448 | 2 | 1 | Embryo 449 | 1 | 1 | Embryo 450 | 2 | 5 |
| Embryo 451 | 5 | 5 | Embryo 452 | 1 | 4 | Embryo 453 | 1 | 3 |
| Embryo 454 | 5 | 5 | Embryo 455 | 5 | 4 | Embryo 456 | 5 | 5 |
| Embryo 457 | 1 | 1 | Embryo 458 | 2 | 5 | Embryo 459 | 1 | 2 |
| Embryo 460 | 2 | 5 | Embryo 461 | 5 | 5 | Embryo 462 | 5 | 5 |
| Embryo 463 | 5 | 2 | Embryo 464 | 5 | 4 | Embryo 465 | 5 | 3 |
| Embryo 466 | 2 | 1 | Embryo 467 | 1 | 1 | Embryo 468 | 5 | 4 |
| Embryo 469 | 5 | 3 | Embryo 470 | 5 | 5 | Embryo 471 | 2 | 4 |
| Embryo 472 | 4 | 4 | Embryo 473 | 3 | 1 | Embryo 474 | 1 | 1 |
| Embryo 475 | 5 | 5 | Embryo 476 | 1 | 2 | Embryo 477 | 1 | 1 |
| Embryo 478 | 3 | 5 | Embryo 479 | 5 | 1 | Embryo 480 | 4 | 5 |
| Embryo 481 | 1 | 2 | Embryo 482 | 5 | 3 | Embryo 483 | 5 | 5 |
| Embryo 484 | 5 | 3 | Embryo 485 | 5 | 2 | Embryo 486 | 3 | 4 |
| Embryo 487 | 5 | 3 | Embryo 488 | 4 | 2 | Embryo 489 | 5 | 2 |
| Embryo 490 | 5 | 3 | Embryo 491 | 5 | 1 | Embryo 492 | 2 | 3 |
| Embryo 493 | 5 | 4 | Embryo 494 | 5 | 3 | Embryo 495 | 2 | 2 |
| Embryo 496 | 2 | 5 | Embryo 497 | 3 | 5 | Embryo 498 | 4 | 5 |

| Embryo 499 | 2 | 3 | Embryo 500 | 5 | 1 | Embryo 501 | 5 | 5 |
|---|---|---|---|---|---|---|---|---|
| Embryo 502 | 1 | 5 | Embryo 503 | 2 | 5 | Embryo 504 | 1 | 1 |
| Embryo 505 | 1 | 5 | Embryo 506 | 1 | 3 | Embryo 507 | 1 | 1 |
| Embryo 508 | 4 | 5 | Embryo 509 | 5 | 5 | Embryo 510 | 2 | 3 |
| Embryo 511 | 1 | 1 | Embryo 512 | 2 | 5 | Embryo 513 | 5 | 5 |
| Embryo 514 | 1 | 5 | Embryo 515 | 3 | 5 | Embryo 516 | 5 | 5 |
| Embryo 517 | 5 | 4 | Embryo 518 | 5 | 1 | Embryo 519 | 5 | 3 |
| Embryo 520 | 5 | 3 | Embryo 521 | 5 | 5 | Embryo 522 | 1 | 1 |
| Embryo 523 | 5 | 3 | Embryo 524 | 5 | 2 | Embryo 525 | 2 | 3 |
| Embryo 526 | 5 | 5 | Embryo 527 | 2 | 1 | Embryo 528 | 4 | 1 |
| Embryo 529 | 5 | 2 | Embryo 530 | 5 | 1 | Embryo 531 | 5 | 2 |
| Embryo 532 | 5 | 3 | Embryo 533 | 5 | 3 | Embryo 534 | 4 | 1 |
| Embryo 535 | 3 | 1 | Embryo 536 | 5 | 2 | Embryo 537 | 3 | 1 |
| Embryo 538 | 5 | 2 | Embryo 539 | 1 | 3 | Embryo 540 | 5 | 3 |
| Embryo 541 | 4 | 1 | Embryo 542 | 3 | 5 | Embryo 543 | 5 | 3 |
| Embryo 544 | 5 | 3 | Embryo 545 | 1 | 1 | Embryo 546 | 1 | 1 |
| Embryo 547 | 4 | 5 | Embryo 548 | 2 | 2 | Embryo 549 | 5 | 5 |
| Embryo 550 | 4 | 5 | Embryo 551 | 1 | 1 | Embryo 552 | 3 | 4 |
| Embryo 553 | 3 | 1 | Embryo 554 | 5 | 4 | Embryo 555 | 1 | 5 |
| Embryo 556 | 2 | 3 | Embryo 557 | 3 | 3 | Embryo 558 | 1 | 3 |
| Embryo 559 | 3 | 4 | Embryo 560 | 3 | 5 | Embryo 561 | 5 | 5 |
| Embryo 562 | 2 | 4 | Embryo 563 | 5 | 5 | Embryo 564 | 2 | 5 |
| Embryo 565 | 3 | 5 | Embryo 566 | 5 | 5 | Embryo 567 | 2 | 3 |

| Embryo 568 | 5 | 5 | Embryo 569 | 5 | 4 | Embryo 570 | 2 | 5 |
|---|---|---|---|---|---|---|---|---|
| Embryo 571 | 2 | 4 | Embryo 572 | 3 | 5 | Embryo 573 | 1 | 5 |
| Embryo 574 | 2 | 3 | Embryo 575 | 5 | 2 | Embryo 576 | 2 | 5 |
| Embryo 577 | 5 | 5 | Embryo 578 | 4 | 4 | Embryo 579 | 1 | 5 |
| Embryo 580 | 5 | 5 | Embryo 581 | 3 | 5 | Embryo 582 | 4 | 1 |
| Embryo 583 | 5 | 2 | Embryo 584 | 3 | 5 | Embryo 585 | 5 | 5 |
| Embryo 586 | 2 | 5 | Embryo 587 | 5 | 3 | Embryo 588 | 5 | 2 |
| Embryo 589 | 5 | 5 | Embryo 590 | 2 | 4 | Embryo 591 | 2 | 5 |
| Embryo 592 | 5 | 1 | Embryo 593 | 4 | 4 | Embryo 594 | 1 | 2 |
| Embryo 595 | 2 | 3 | Embryo 596 | 3 | 5 | Embryo 597 | 2 | 4 |
| Embryo 598 | 5 | 5 | Embryo 599 | 5 | 5 | Embryo 600 | 5 | 3 |
| Embryo 601 | 5 | 5 | Embryo 602 | 4 | 5 | Embryo 603 | 5 | 5 |
| Embryo 604 | 5 | 5 | Embryo 605 | 5 | 4 | Embryo 606 | 5 | 5 |
| Embryo 607 | 5 | 1 | Embryo 608 | 5 | 5 | Embryo 609 | 5 | 3 |
| Embryo 610 | 5 | 3 | Embryo 611 | 5 | 5 | Embryo 612 | 5 | 5 |
| Embryo 613 | 5 | 5 | Embryo 614 | 4 | 5 | Embryo 615 | 5 | 5 |
| Embryo 616 | 3 | 5 | Embryo 617 | 5 | 5 | Embryo 618 | 1 | 5 |
| Embryo 619 | 1 | 1 | Embryo 620 | 5 | 5 | Embryo 621 | 3 | 1 |
| Embryo 622 | 5 | 5 | Embryo 623 | 5 | 2 | Embryo 624 | 5 | 4 |
| Embryo 625 | 1 | 1 | Embryo 626 | 5 | 5 | Embryo 627 | 5 | 4 |
| Embryo 628 | 5 | 5 | Embryo 629 | 5 | 4 | Embryo 630 | 5 | 5 |
| Embryo 631 | 5 | 5 | Embryo 632 | 5 | 4 | Embryo 633 | 5 | 5 |
| Embryo 634 | 5 | 5 | Embryo 635 | 5 | 5 | Embryo 636 | 5 | 5 |
| Embryo 637 | 3 | 1 | Embryo 638 | 3 | 5 | Embryo 639 | 5 | 5 |
| Embryo 640 | 1 | 4 | Embryo 641 | 1 | 5 | Embryo 642 | 1 | 3 |
| Embryo 643 | 2 | 1 | Embryo 644 | 3 | 5 | Embryo 645 | 5 | 3 |
| Embryo 646 | 1 | 5 | Embryo 647 | 4 | 5 | Embryo 648 | 3 | 4 |
| Embryo 649 | 4 | 3 | Embryo 650 | 2 | 1 | Embryo 651 | 3 | 4 |
| Embryo 652 | 5 | 5 | Embryo 653 | 5 | 5 | Embryo 654 | 5 | 5 |
| Embryo 655 | 5 | 1 | Embryo 656 | 1 | 2 | Embryo 657 | 5 | 2 |
| Embryo 658 | 3 | 1 | Embryo 659 | 2 | 5 | Embryo 660 | 1 | 1 |
| Embryo 661 | 5 | 5 | Embryo 662 | 1 | 1 | Embryo 663 | 4 | 1 |
| Embryo 664 | 5 | 1 | Embryo 665 | 5 | 5 | Embryo 666 | 1 | 3 |
| Embryo 667 | 5 | 5 | Embryo 668 | 2 | 2 | Embryo 669 | 4 | 3 |
| Embryo 670 | 4 | 3 | Embryo 671 | 5 | 5 | Embryo 672 | 5 | 3 |
| Embryo 673 | 5 | 2 | Embryo 674 | 1 | 5 | Embryo 675 | 3 | 5 |
| Embryo 676 | 5 | 5 | Embryo 677 | 2 | 4 | Embryo 678 | 5 | 5 |
| Embryo 679 | 3 | 5 | Embryo 680 | 5 | 5 | Embryo 681 | 3 | 5 |
| Embryo 682 | 1 | 2 | Embryo 683 | 3 | 3 | Embryo 684 | 1 | 4 |
| Embryo 685 | 5 | 5 | Embryo 686 | 5 | 5 | Embryo 687 | 3 | 2 |
| Embryo 688 | 5 | 3 | Embryo 689 | 3 | 5 | Embryo 690 | 3 | 4 |
| Embryo 691 | 2 | 5 | Embryo 692 | 3 | 1 | Embryo 693 | 3 | 4 |
| Embryo 694 | 1 | 1 | Embryo 695 | 3 | 5 | Embryo 696 | 3 | 5 |
| Embryo 697 | 5 | 5 | Embryo 698 | 2 | 5 | Embryo 699 | 1 | 5 |
| Embryo 700 | 5 | 1 | Embryo 701 | 5 | 4 | Embryo 702 | 5 | 4 |
| Embryo 703 | 5 | 5 | Embryo 704 | 3 | 1 | Embryo 705 | 5 | 3 |

| Embryo 706 | 5 | 5 | Embryo 707 | 5 | 5 | Embryo 708 | 2 | 5 |
|---|---|---|---|---|---|---|---|---|
| Embryo 709 | 5 | 5 | Embryo 710 | 1 | 1 | Embryo 711 | 2 | 5 |
| Embryo 712 | 3 | 5 | Embryo 713 | 3 | 1 | Embryo 714 | 5 | 3 |
| Embryo 715 | 3 | 4 | Embryo 716 | 3 | 1 | Embryo 717 | 5 | 5 |
| Embryo 718 | 3 | 5 | Embryo 719 | 2 | 5 | Embryo 720 | 3 | 5 |
| Embryo 721 | 3 | 5 | Embryo 722 | 1 | 1 | Embryo 723 | 2 | 5 |
| Embryo 724 | 3 | 5 | Embryo 725 | 3 | 1 | Embryo 726 | 1 | 3 |
| Embryo 727 | 3 | 4 | Embryo 728 | 2 | 5 | Embryo 729 | 4 | 5 |
| Embryo 730 | 5 | 5 | Embryo 731 | 5 | 1 | Embryo 732 | 5 | 5 |
| Embryo 733 | 1 | 1 | Embryo 734 | 4 | 5 | Embryo 735 | 5 | 1 |
| Embryo 736 | 4 | 3 | Embryo 737 | 3 | 5 | Embryo 738 | 5 | 5 |
| Embryo 739 | 1 | 1 | Embryo 740 | 1 | 1 | Embryo 741 | 1 | 1 |
| Embryo 742 | 1 | 1 | Embryo 743 | 1 | 1 | Embryo 744 | 1 | 1 |
| Embryo 745 | 3 | 4 | Embryo 746 | 1 | 1 | Embryo 747 | 2 | 1 |
| Embryo 748 | 1 | 1 | | | | | | |

**Table S3: Raw data of the system's classifications with the actual annotations for 70 hpi evaluations using the test set.** Blue highlights indicate positive or blastocyst development at 113 hpi and red highlights indicate negative or non-blastocyst development at 113 hpi. Class values range from 1-5 and represent the training classes 1-5.